\documentclass[useAMS,usenatbib]{mn2e}
\usepackage{natbib}
\usepackage{color,graphicx} 
\usepackage{amsmath}        
\usepackage{amsfonts}       
\usepackage{amssymb}        
\usepackage{hyperref} 
\usepackage{wasysym}
\title{ULTRACAM \textit{z'}-band Detection of the Secondary Eclipse of WASP-12b}
\author[D F\"{o}hring et al.]{D. F\"{o}hring,$^{1}$\thanks{E-mail: dora.fohring@durham.ac.uk}
V. S. Dhillon,$^{2}$ Nikku Madhusudhan,$^{3}$ T. R. Marsh,$^{4}$ 
\newauthor 
C. M. Copperwheat,$^{5}$  S. P. Littlefair,$^{2}$ Richard W. Wilson$^{1}$ \\
$^{1}$Department of Physics, Centre for Advanced Instrumentation, University of Durham, South Road, Durham, DH1 3LE, UK \\
$^{2}$Department of Physics and Astronomy, University of Sheffield, Sheffield S3 7RH, UK \\
$^{3}$Department of Physics and Department of Astronomy, Yale University, New Haven, CT 0651, USA \\
$^{4}$Department of Physics, University of Warwick, Coventry CV4 7AL, UK \\
$^{5}$Astrophysics Research Institute, Liverpool John Moores University, Wirral, CH41 1LD, UK }

\begin{document}
\maketitle

\begin{abstract}

We present \textit{z'}-band secondary eclipse photometry of the highly irradiated hot Jupiter WASP-12b using ULTRACAM on the 4.2m William Herschel Telescope. We measure a decrease in flux of $\delta=0.130 \pm0.013 \%$ during the passage of the planet behind the star, which is significantly deeper than the previous measurement at this wavelength ($0.082 \pm0.015 \%$, \citealt{morales2010}).
Our secondary eclipse is best fit with a mid-eclipse phase, $\phi$, that is compatible with a circular orbit $\phi=0.501 \pm0.002$, in agreement with previous results \citep{croll2011}. 
In combination with existing data, our eclipse depth measurement allows us to constrain the characteristics of the planet's atmosphere, which is consistent with a carbon-rich model, with no evidence for a strong thermal inversion. If the difference in eclipse depth reported here compared to that of \citet{morales2010} is of physical origin, as opposed to due to systematics, it may be caused by temporal variability in the flux, due to atmospheric dynamics.
\end{abstract}

\begin{keywords}
planetary systems -- stars: individual: WASP-12 -- techniques: photometric
\end{keywords}


\section{Introduction} \label{intro}

Observations of secondary eclipses provide a powerful means of obtaining information about exoplanetary systems. During a secondary eclipse, as the planet passes behind its host star, a decrease in flux directly corresponding to the planet's emission is observed. Photometric measurements of the secondary eclipse allow us to probe the planet's thermal structure \citep{deming2005}, atmospheric composition \citep{deming2006} and constrain the orbital eccentricity \citep{charbonneau2005}, all which are key in assessing the planet's albedo, energy budget and tidal history. To date, ground-based observations have proved to be invaluable in aiding the atmospheric characterisation of transiting hot Jupiters (e.g. \citealt{sing2009, demooij2009, croll2011, zhao2012}).

The transiting exoplanet WASP-12b is a large (1.79 $R_J$) hot Jupiter on a close (0.0229 AU, 1.09 day) orbit around a 6300 K host star \citep{hebb2009}. It is one of the largest and most intensely irradiated planets currently known, with an equilibrium temperature of $T_{eq}$ = 2516 K. The combination of a relatively bright (1.35 $M_{\astrosun}$, G0V) host star and a large close-in planet with an orientation which enables the viewing of both primary and secondary events make WASP-12b an ideal system for ground-based follow-up observations. 

The overinflated radius of WASP-12b poses a challenge for standard planetary models, which predict an upper limit of 1.2$R_J$ for evolved gas giants \citep{bodenheimer2003}. 
Proposed mechanisms for inflation include tidal heating \citep{ibgui2010}, Ohmic dissipation (e.g. \citealt{pernahengpont2012, batyagin&stevenson2010, huang&cumming2012}), or strong irradiation (e.g. \citealt{enoch2012}).
Previous studies of WASP-12b have suggested intense tidal forces, giving it a  prolate shape and causing it to lose mass to its host star through tidal stripping \citep{li2010}. There is also evidence for the presence of a bow shock around the planetary magnetosphere directly in front of the planet \citep{llama2011}. Previous spectroscopic studies have found that the atmosphere of WASP-12b is lacking a strong thermal inversion and possesses very efficient day-night energy circulation, as well as an extreme overabundance of carbon, both of which are contrary to previous theoretical predictions \citep{madhusudhan2011}, although this has been recently questioned by newer photometric data \citep{crossfield2012}.  

Theory predicts that the timescale for circularization of WASP-12b is very short. Initial radial velocity and eclipse measurements have suggested an eccentric orbit \citep{hebb2009, morales2010}, but this was later refuted \citep{croll2011, campo2011, husnoo2011}. An eccentric orbit would indicate orbital pumping from one or more undetected planets, or a very low tidal dissipation \citep{li2010}.

Previous observations of emission from the dayside atmosphere of WASP-12b have been reported using the Spitzer Space Telescope, at 3.6 $\mu$m, 4.5 $\mu$m, 5.8 $\mu$m and 8 $\mu$m  \citep{campo2011}, and from ground-based observations in the \textit{J}, \textit{H}, \textit{Ks}  \citep{croll2011, zhao2012} and \textit{z'} bands \citep{morales2010}. 

In this paper, we present a new, high-precision detection of the \textit{z'} (9097\AA) emission from the extrasolar planet WASP-12b, and discuss the implications on our understanding of the planet. 
In section \ref{obsan} we describe our observations and data reduction. In section \ref{em} we detail our light curve modelling and eclipse detection. In section \ref{ee} we show our method of error estimation, and in section \ref{atmos}, we compare the emission of the planet to atmospheric models. Finally, we discuss our results and provide conclusions in sections \ref{disc} and \ref{conc}.

\section{Observations and Data Reduction} \label{obsan}

The secondary eclipse of WASP-12b was observed on 2010 January 5 - 6 at the 4.2m William Herschel Telescope on La Palma using ULTRACAM \citep{dhillon2007}.  
A total of 3592 observations were made in the SDSS \textit{u'}, \textit{g'}, and  \textit{z'} filters between 22:54 UT and 05:30 UT during excellent transparency conditions. 
We obtained 3.05 hours of out-of-eclipse data and 2.70 hours of in-eclipse data at airmasses between 1.000 - 2.319. 
The telescope was defocussed to give a stellar FWHM of $\sim$5", enabling longer integration times without saturation, and reducing the effects of pixel sensitivity variations across the stellar profile (see \citealt{southworth2009}). An exposure time of 5.6s was used throughout the run, with a dead time between each exposure of 25ms.  
Each frame was timestamped to sub-millisecond accuracy using a dedicated GPS system \citep{dhillon2007}.

The data were corrected for bias and flat-field using the ULTRACAM pipeline software\footnote{\url{http://deneb.astro.warwick.ac.uk/phsaap/software/ultracam/html/index.html}}. The effect of correcting the fringing that was apparent in the \textit{z'} filter was investigated in detail using a fringe-frame constructed from night-sky frames;  it was found that this correction did not produce a marked difference in the light curves and so was not employed in the final reduction. 

We performed aperture photometry on the target and the brightest 9 comparison stars in the $5' \times 5'$ field using aperture sizes of radii 15 - 50 pixels (4.5 - 15 arcseconds). The best aperture was the one that produced the most stable photometry with the lowest variance in segments of the out-of-eclipse and mid-eclipse regions of the resulting light curve, and was  found using a golden section search \citep{kiefer1953}. For our data this resulted in an aperture of 42 pixels. The centres of each star were determined by fitting a Moffat profile \citep{moffat1969} to the PSF of each star. We decided to add an extra aperture on our target to include a background star, $\sim$10" from WASP-12 and partially overlapping our optimum aperture, as this improved the variance. The implications of this are discussed in section \ref{ee}. We found that the size of the sky annulus did not have a significant effect on the photometry, and kept it constant with an inner radius of 60 pixels and outer radius of 90 pixels. The sky background level was determined using a clipped mean method, with fainter background stars masked out from the sky annuli. 

We produced a single light curve by combining individual differential light curves that were produced from the three brightest comparison stars in the field, which had similar flux levels to the target in the \textit{z'} band. The individual light curves were weighted based on the inverse chi-square of a straight-line fit to their out-of-eclipse and mid-eclipse sections. The scatter per point on this final light curve was measured as $0.14$ mmag, or 0.013\%.
 
In our data, the total drift in the pointing of WASP-12 on the CCDs was 14 pixels in \textit{x} and 24 pixels in \textit{y}.
We tested our data for systematic correlations with \textit{x} and \textit{y} positions of the target on the detector, seeing, sky brightness variations and telescope parallactic angle. This was done by fitting linear correlations between each parameter and the out-of-eclipse portions of the light curve. We did not find any correlations. We discarded the latter 1.36 hours of the out-of-eclipse data precisely at the onset of sudden transparency variations, as they caused the scatter per point to increase by up to a factor of 100. This left a total of 2740 observations between 22:54 UT and 04:08 UT, covering phases 0.426 and 0.594 based on the ephemeris of \citet{hebb2009}, and air masses between 1.000 - 1.477. 
\begin{figure}
   \centering
   \includegraphics[width=3.5in]{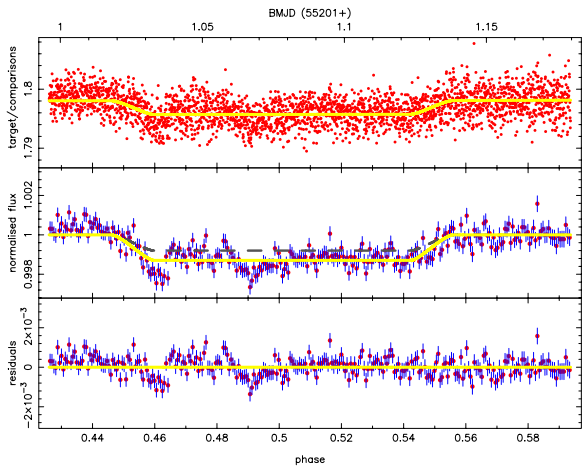}
 \caption{Secondary eclipse of WASP-12b observed in the \textit{z'} band. The top panel shows the unbinned raw light curve, with the best-fit secondary eclipse model (including background) overplotted in yellow (see section \ref{em}). The middle panel shows the background subtracted and normalised light curve with the data binned into groups of 10 data points, corresponding to 1 minute, again with the best-fit model overplotted. The grey dashed line shows a depth of 0.082\% as found by \citet{morales2010}. The bottom panel shows the residuals after subtraction of the best-fit model from the data shown in the middle panel.}
   \label{lc}
\end{figure}
\section{Eclipse modelling} \label{em} 

To establish the presence of an eclipse we modelled the light curve using the technique outlined by \citet{sackett1999} without limb darkening. The input parameters of the model include the radii of the star and planet, the semi-major axis, the orbital inclination and the ephemeris, as given by \citet{hebb2009}. We also attempted to use the system parameters determined by \citet{chan2011}, but found that the duration of the eclipse matches that found by \citet{hebb2009} more closely and yields a lower chi-squared fit. 

The best-fit model was found by using a downhill simplex $\chi^2$ minimisation routine \citep{press2007}, with the eclipse depth, $\delta$, the phase of mid-eclipse, $\phi$, and two background terms, $c_1$ and $c_2$, as free parameters. The two background terms allow the out-of-eclipse differential flux, $B$, to be fitted by an equation of the form: $B = c_1 + c_2 X$, where $X$ is the airmass. The best-fit model, shown in Figure \ref{lc}, is an eclipse with a depth of $\delta = 0.130\%$, centered at an orbital phase $\phi = 0.501$. The reduced $\chi^2$ of the fit is 1.083. We also searched for evidence of a secondary eclipse in the $g'$ and $u'$ bands and found our results consistent with no detection in either.

We performed two tests to confirm the eclipse detection, in a manner similar to previously reported eclipse results (\citealt{deming2005, rogers2009, morales2010}).
In the first test, we averaged all of the fully in-eclipse data points from the background-corrected light curve and repeated this for all of the out-of-eclipse points. The difference between the two measurements gives an eclipse depth of $0.135\% \pm  0.005\%$, where the error is calculated from the scatter on our in-eclipse and out-of-eclipse data points. 

\begin{figure}
   \centering
   \includegraphics[width=2.5in]{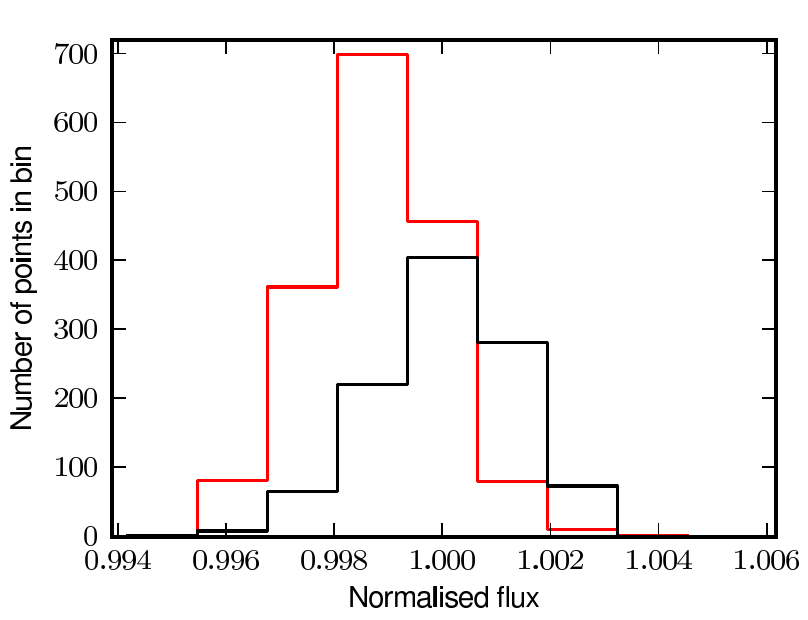} 
   \caption{Histograms of the in-eclipse (red line) and out-of-eclipse portions of the normalised WASP-12b light curve. The width of each bin is $0.13\%$, the same as the eclipse depth.}
    \label{hist}
\end{figure}

In the second test,  we generated histograms of the in-eclipse and out-of-eclipse portions of the normalised  light curve, and set the bin width equal to the measured eclipse depth.  The result, shown in Figure \ref{hist}, illustrates how the flux distribution of in-eclipse points is shifted by one bin, as expected, with respect to the out-of-eclipse flux distribution centred on 1.0.

\section{Error Estimation} \label{ee}
To compute the error in the eclipse depth and to determine the extent of systematics, we employ the binning technique described by \citet{pont2006}. 
Using this method, the uncertainty on the eclipse depth, $\sigma(N)$, is approximated as the error on the mean of the in-eclipse flux. In the absence of correlated noise, $\sigma(N) = \sigma_w /\sqrt{N}$, where N is the number of points in a given interval, here the eclipse, and $\sigma_w$ is the measurement uncertainty, which can be obtained from the uncorrelated rms scatter per out-of-eclipse data points. When taking into account the covariance of the data on the timescales of the eclipse, the error on the eclipse depth can be shown to be well modelled by the relation  $\sigma(N)^2	= \sigma_w^2 /N+ \sigma_r^2$, where $\sigma_r$ is the systematic (red) noise component. The uncertainty $\sigma(N)$ can be estimated from the data by binning the out-of-eclipse points in time in intervals of $N$ points: the root mean square scatter of the binned residuals will follow an $N^{-1/2}$ relation in the absence of red noise; otherwise, they will be fit by $\sigma^2(N)	= \sigma_w^2 /N + \sigma_r^2$.
Binning our data up to the timescales of the ingress and egress duration of 21 minutes (N = 220), we find $\sigma(220) \simeq \sigma_r = 0.013\%$.

\begin{figure}
   \centering
   \includegraphics[width=3.3in]{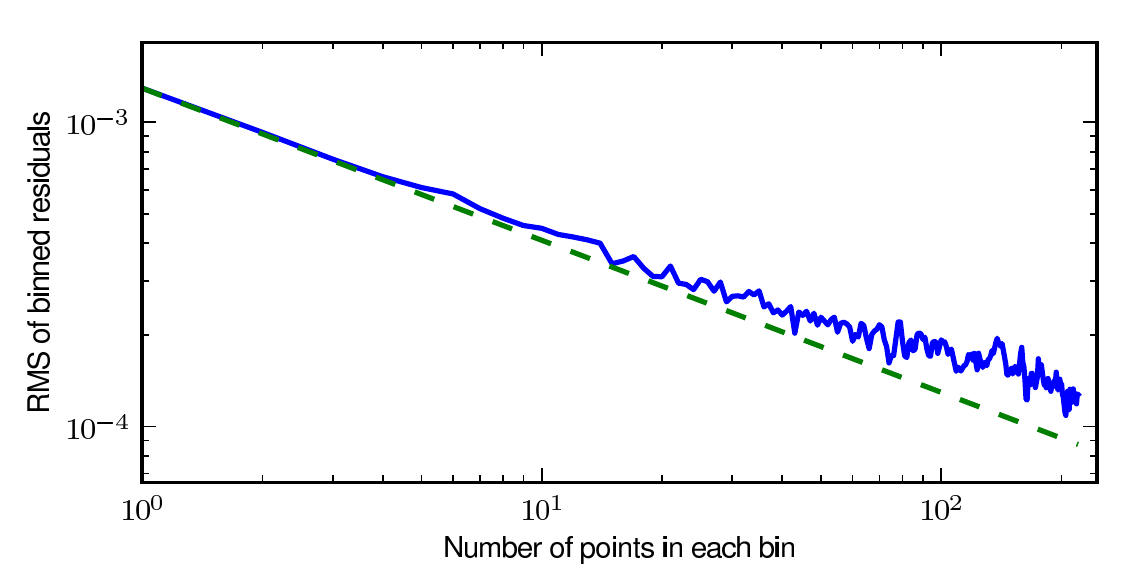} 
   \caption{The root-mean-square scatter of our out-of-eclipse photometry (solid line), plotted as a function of the light curve binning factor. The dashed line displays the expectation for Gaussian noise.}
      \label{bi}
\end{figure}

We investigated the extent to which uncertainties in the system parameters affect our eclipse depth and mid-eclipse results.
By varying the inclination, planet-to-star radius ratio, and semi-major axis of the system by $1 \sigma$ from the reported values in \citet{hebb2009}, the measured eclipse depth changed by a maximum of $3.59\%$ or $0.28 \sigma$, while the mid-eclipse phase changes by $0.0004 \%$. 

We calculated two independent values of the errors in the eclipse depth and the phase of mid eclipse using the prayer-bead \citep{gillon2009} and bootstrapping \citep{press2007} techniques in order to estimate the error on the mid-eclipse phase. These provided estimates of $\phi=0.501 \pm  0.002$, $\delta = 0.130 \%\pm  0.002\%$ and $\phi=0.5014 \pm  0.0004$, $\delta = 0.1298\% \pm  0.0003\%$, respectively. 
Taking the final values for eclipse depth and mid eclipse phase from the downhill simplex results and the errors from the most conservative estimate out of the \citet{pont2006} method, prayer-bead and bootstrap yielded the final result for the phase of mid eclipse and the eclipse depth as $\phi = 0.501 \pm0.002$ and $\delta = 0.130 \pm0.013 \%$. 

It is necessary to discuss the implication of including additional background stars in our aperture photometry, as discussed in section \ref{obsan}.
In their paper, \citet{crossfield2012} calculate that including an unresolved M dwarf star with a 1" separation from WASP-12 (\citealt{bergfors2011, bergfors2013}) in the photometry aperture dilutes the measured eclipse depth by $3.97\%$ in the \textit{z'} band. Accounting for this star in our measurement results in an eclipse depth of 0.135\%. A second faint background star was also included in our aperture, which we have determined from our in-focus images contributes an additional dilution of 0.9\%. This gives a total corrected eclipse depth of of 0.136\%, which is within our error of $\pm0.013\%$.

\section{Comparison with Atmospheric Models} 
\label{atmos}

\begin{figure}
   \centering
   \includegraphics[width=3.5in]{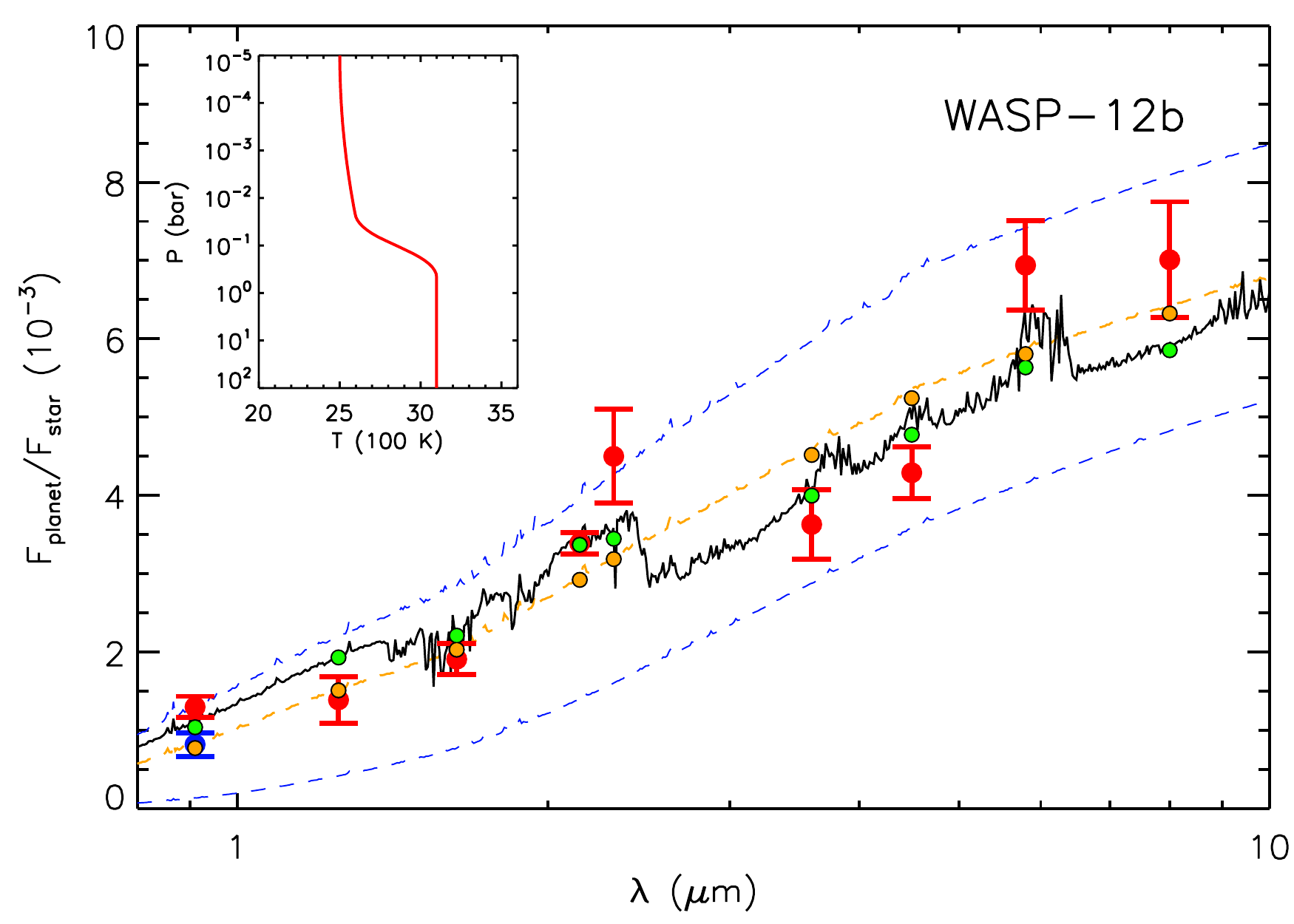}    
   \caption{Observations and model secondary eclipse spectrum of WASP-12b. Our \textit{z'}-band eclipse depth is shown as the red filled circle with error bars at 0.91 $\mu$m, together with previously published photometric observations at other wavelengths reported by \citet{croll2011}, \citet{campo2011}, \citet{cowan2012} and \citet{crossfield2012}. The blue circle with error bars beneath our data shows the previously published data in \textit{z'} band by \citet{morales2010}  (see Section~\ref{atmos}).  The black solid curve represents a best-fit model emission spectrum of the dayside atmosphere of WASP-12b with a carbon-rich composition and no thermal inversion. The orange dashed curve shows the best fitting blackbody model with a temperature of 3135 K. The green circles without error bars represent the carbon-rich model integrated over the bandpasses of the data, while the orange circles are the integrated points for the blackbody model. The temperature profile is shown in the inset. The blue dashed curves are two blackbody spectra with temperatures of 2200 K and 3200 K shown for reference. The current data are best fit by a carbon-rich model without a strong thermal inversion, consistent with the findings of \citet{madhusudhan2011}.} 
\label{fig:spectrum}
\end{figure}

We used our \textit{z'}-band planet-star flux contrast, together with previously published data, to constrain the atmospheric properties of WASP-12b. We modelled the dayside atmosphere of WASP-12b using the approach followed in \citet{madhusudhan2012}. The model, first developed in  \citet{madhusudhanseager2009}, considers 1-D line-by-line radiative transfer in a plane-parallel atmosphere, with constraints of local thermodynamic equilibrium, hydrostatic equilibrium, and global energy balance, and includes the major molecular and collision-induced opacity sources expected in hydrogen-rich atmospheres. The temperature profile and molecular abundances constitute 12  free parameters of the model; six parameters for the temperature profile and six for the molecular abundances of H$_2$O, CO, CH$_4$, CO$_2$, C$_2$H$_2$, and HCN. Given the data, we explored the space of temperature profiles and composition, both oxygen-rich and carbon-rich \citep{madhusudhan2012}, using a Markov Chain Monte Carlo optimisation algorithm \citep{madhusudhan2011} to find model fits to data.  two values have been re

We assumed that all the \textit{z'}-band flux arose from thermal emission, in accordance with \citet{morales&seager2007}, who showed that reflected light emission contributes 10-20 times less than thermal emission for very hot Jupiters, such as WASP-12b. We fitted models of WASP-12b to our observed \textit{z'}-band flux together with existing photometric data in the $J$, $H$, and $K$ bands \citep{croll2011}, four-channels of \textit{Spitzer} IRAC photometry \citep{campo2011,cowan2012}, and a narrow-band photometric observation at 2.3 $\mu$m \citep{crossfield2012}. For the two IRAC data points at 3.6 and 4.5 $\mu$m, we adopt the values from \citet{cowan2012}, which have more conservative error bars. For the 4.5 $\mu$m point, for which two values have been reported by \citet{cowan2012}, we choose the value that is consistent with the value reported by \citet{campo2011}. For all the previously published data, we use the revised values from \citet{crossfield2012} which have been corrected for the presence of a stellar companion. A model fit to all the data is shown in Figure~\ref{fig:spectrum}. We found that a carbon-rich model (with C/O $\ge$ 1) and a weak thermal inversion in its dayside atmosphere provides the best fit to all the present data, in agreement with \citet{madhusudhan2011} and \citet{madhusudhan2012}. The model spectrum shown in Figure~\ref{fig:spectrum}, of an atmosphere that is carbon-rich and has no thermal inversion, provides a good fit to the data and has a $\chi^2$ of 22.9. We also explored models with isothermal temperature profiles as suggested by \citet{crossfield2012}, and found that the best fitting isothermal model (with one free parameter) has a temperature of 3135 K and a $\chi^2$ of 48.8. To account for the different number of free parameters, we compared the models using the Bayesian Information Criterion \citep{schwarz1978}, defined as BIC =Ê$\chi^2$ + $N_{par}$ $\times$ ln($N_{dat}$), where $N_{par}$ is the number of free parameters, and $N_{dat}$ is the number of data. The BIC for the carbon-rich model was 49.3, lower than the value of 51.0 obtained for the isothermal model, indicating a better fit. An oxygen-rich model with or without a thermal inversion also provides a worse fit to the data compared to the carbon-rich model.An oxygen-rich model with or without a thermal inversion also provides a worse fit to the data compared to the carbon-rich model. Consequently, the present data favour the interpretation of a carbon-rich atmosphere with no thermal inversion in WASP-12b. 

A more detailed atmospheric retrieval analysis and new data, which are beyond the scope of the current work, would be required to place tighter constraints on the C/O ratio of WASP-12b. In particular, we have used the {\it Spitzer} IRAC 3.6 and 4.5 $\mu$m flux contrasts reported by \citet{cowan2012} and our present observation in the z' band. In doing so, a few other combinations of previous datasets at the same wavelengths have not been considered here. For example, considering the {\it Spitzer} IRAC 3.6 and 4.5 $\mu$m data from \citet{campo2011}, which have lower uncertainties, would make the carbon-rich model more likely and the isothermal model even less likely. On the other hand, we have used only one of the two values for the IRAC 4.5 $\mu$m point reported by \citet{cowan2012} that is consistent with the value of \citet{campo2011}. Considering the alternate point could improve the fit for the isothermal model. Yet other combinations result from using the previously reported \textit{z'}-band flux of \citet{morales2010} instead of our current observation in the same band. We also note that a high-confidence narrow-band photometric observation at 2.3 $\mu$m might be able to provide a good constraint on the CO abundance. Future observations at this wavelength with more out-of-eclipse data than reported in \citet{crossfield2012} could provide a longer temporal baseline for a reliable estimate of the eclipse depth. We have also not used the thermal spectrum of WASP-12b obtained in the HST WFC3 bandpass (1.1 - 1.6 $\mu$m), reported by \citet{swain2012}, as it has not yet been corrected for contamination due to the stellar companion \citep{crossfield2012}. A detailed atmospheric retrieval analysis of WASP-12b based on the various combinations of these existing, and possibly re-reduced and/or corrected, datasets will be pursued in a future study. 

\section{Discussion} \label{disc}

Our observed eclipse depth, i.e. the planet-star flux ratio, in the \textit{z'} band is 58\% higher than that reported by \citet{morales2010}. This disagreement is at the 2.4-$\sigma$ level and there are examples in the literature suggesting that this is not uncommon, e.g. the $z'$-band measurements of WASP-19b by \citet{burton2012} and \citet{lendl2013}. If this difference is of physical origin, as opposed to a systematic error, it may be evidence for temporal variability in the stellar or planetary flux. In principle, stellar variability may be caused due to stellar flares, as suggested in \citet{haswell2012}, although this is unlikely given the spectral type of the host star. On the other hand, if the variability is in the emission from the planet it would imply temporal changes in the temperature distribution on the dayside hemisphere of WASP-12b at pressures close to the planetary photosphere in 
the \textit{z'} band, typically around 1 bar.

Assuming the stellar flux in the \textit{z'} band is constant, our observation implies a 58\% increase in the hemispherically-averaged dayside thermal emission from WASP-12b over that observed by \citet{morales2010}. Since the two observations were separated in time by 2 months, it may be possible that transient effects due to atmospheric dynamics might be causing the observed differences. Indeed, some general circulation models (GCMs) of hot Jupiters predict that large enough spatial structures, e.g. circumpolar vortices, in their atmospheres could cause temporal photometric variations in thermal emission from their dayside atmospheres (e.g. \citealt{rauscher2007, cho2008}; but cf. \citealt{showman2009atmospheric}). \citet{rauscher2007} explored possible photometric variability in several hot Jupiters as a function of their mean zonal wind speeds and radiative forcing. For the example of hot Jupiter HD 209458b, their models showed variations as high as 60-70\% for wind speeds of 800 m/s. While a similar calculation has not been reported in the literature for WASP-12b, the order of magnitude higher incident irradiation received by WASP-12b compared to HD 209458b would imply higher wind speeds in WASP-12b and might cause the observed variability. GCM models of WASP-12b in the future would be able to investigate this possibility. Recent work by \citet{fossati2013} has shown that variability of a G-type star due to star spots would cause a  magnitude variation of $\sim 0.2\%$. This amount is too low to explain our variable eclipse depth. Other work by \citet{henry2013} has found absorbing gas around WASP-12. A variable cloud of gas might cause some Rayleigh scattering, but that would be significant mainly in the visible, and not much in the red-optical where our \textit{z'} band lies. The effect of a steady-state circumstellar disk or cloud of gas should cancel out when we subtract the in-eclipse and out-of-eclipse fluxes to obtain the eclipse depth. 

We have also explored the possibility of local variations in the surface brightness of the dayside atmosphere of WASP-12b caused due to potential transient effects, e.g. ``storms", following \citet{agol2010climate}. We considered a two-component \textit{z'}-band brightness temperature distribution of the dayside hemisphere of the planet - a homogeneous background temperature ($T_0$) and a local perturbation with a temperature $T_1 = T_0 + \Delta T$ over a region with a parametric covering fraction ($f$), i.e. the fraction of the surface area covered by the perturbation. Assuming nominal covering fractions  of $f = 0.1 - 0.2$ (\citealt{showman2009atmospheric, agol2010climate}), we find that the difference between the two observed \textit{z'}-band fluxes can be caused by temperature fluctuations of $\sim 450 - 750$ K or higher, relative to a background temperature of $\sim 3000$ K. 

\section{Conclusions} \label{conc}

We have detected emission from WASP-12b in the \textit{z'} band. We measure the eclipse depth at 0.91$\mu$m to be $0.130 \pm0.013$ per cent, significantly deeper in comparison to previous data. Assuming this difference is not caused by systematics, and that the stellar flux is unchanging, given the early spectral type of the star, the observed discrepancy may be caused by a temporally variable photospheric temperature of the planet. Local thermal fluctuations in the surface brightness distribution of the planet caused by atmospheric dynamics may be able to explain the observed variability, which can be further constrained using general circulation models of WASP-12b in the future. Considering our  \textit{z'}-band observation along with previously reported photometric observations in other bandpasses, we find that the sum-total of data are best explained by a carbon-rich model, with no evidence for a strong thermal inversion, as has been previously reported. We also estimate the mid-eclipse phase of the planet to be $0.501 \pm0.002$, which corresponds to a circular orbit. 
 
\section{Acknowledgements} \label{ack}
DF, VSD, TRM, SPL, CMC \& ULTRACAM are supported by the STFC. Based on observations obtained with the 4.2m William Herschel Telescope operated on the island of La Palma by the Isaac Newton Group in the Spanish Observatorio del Roque de los Muchachos of the Instituto de Astrofsica de Canarias. NM acknowledges support from the Yale Center for Astronomy and Astrophysics (YCAA) through the YCAA postdoctoral prize fellowship at Yale University.

\bibliographystyle{mn2e}
\bibliography{bib}

\end{document}